# Collective responses of Bi-2212 stacked junction to 100 GHz microwave radiation under magnetic field oriented along the *c*-axis

V.N. Pavlenko[1], Yu.I. Latyshev[1*], J. Chen[2], M.B. Gaifullin[3], A. Irzhak[4], S.-J. Kim[5], and P.H. Wu[2]

[1]*Institute of Radio-Engineering and Electronics RAS, Mokhovaya 11/7, 125009 Moscow, Russia*[4]

[2]*Nanjing University, RISE, 22 Hankou Road, Nanjing 210093, China*

[3]*Loughborough University, Loughborough LE11 3TU, United Kingdom*

[4]*Moscow Institute of Steel and Alloys, Leninsky prospect 4, 119049 Moscow, Russia*

[5]*Cheju National University, 1 Ara 1-dong, Jeju, Jeju-do, Korea 690-756*

*Corresponding author*: e-mail: yurilatyshev@yahoo.com

**Abstract.** We studied a response of Bi-2212 mesa type structures to 100 GHz microwave radiation. We found that applying magnetic field of about 0.1 T across the layers enables to observe collective Shapiro step response corresponding to a synchronization of all 50 intrinsic Josephson junctions (IJJ) of the mesa. At high microwave power we observed up to 10[th] harmonics of the fundamental Shapiro step. Besides, we found microwave induced flux-flow step position of which is proportional to the square root of microwave power and that can exceed at high enough powers 1 THz operating frequency of IJJ oscillations.

Layered high-temperature superconducting materials, such as $Bi_2Sr_2CaCu_2O_8$ (Bi-2212), are composed of superconducting $CuO_2$ double layers coupled by a Josephson interaction. This unique system exhibits the Josephson effects at the atomic scale. The discovery of intrinsic Josephson effects in these superconductors [1] opens up a new nanoscale route for developing superconducting electronics. For many applications the important requirement is to provide synchronization of Josephson oscillations in elementary junctions of the stack. This synchronization can be probed by Shapiro step response of the stack to the external microwaves since the voltage position of collective response indicates the number of synchronized junctions.

The first experiments on THz response of Bi-2212 stacks have shown that IJJs can operate up to frequencies of 2.5 THz [2], however, they are not effectively coupled by electric field across the stack due to the screening effects [3]. As a result, each elementary IJJ responds to the microwaves individually [2].

More effective coupling of IJJs can be provided by magnetic field parallel to the layers [4]. Collective Shapiro step of 60 IJJs synchronized by parallel magnetic field has been observed in subterahertz frequency range [5]. However, to operate at 150 GHz one needs to apply quite high parallel field of about 2.5 T.

In the present paper we demonstrate the alternative way to synchronize IJJs to subterahertz radiation using much lower magnetic field of the order of 0.1 T oriented across the layers.

The structures have been fabricated by double sided processing of thin Bi-2212 single crystal whiskers by focused ion beam. We simplified the method described in Ref. [6]. The stages of fabrication are shown in Fig. 1. First, the square groove with the depth of half of the crystal thickness $d$ is etched and marked at the corners (Fig. 1a). The depth is controlled by one half etching time of the same area to the full thickness. Second, a crystal is turned over and another square groove is made on the other side of crystal near the first one, aligned against the markers (Fig.1b). The depth of the second groove slightly exceeds $d/2$. At the third stage the structure is trimmed as shown in Fig. 1c and the junction is formed between cutting lines. Typical structure had lateral sizes of about 1μm x 1μm and contained 20-100 JJs along the *c*-axis. Before the second processing the crystal has been glued to the MgO substrate by solution of collodion in amyl acetate. A collodion then has been removed from the top of the crystal by oxygen discharge. The four golden contact pads have been evaporated and annealed before FIB processing to avoid diffusion of Ga-ions into the mesa. The substrate with a mesa has been mounted in the optical cryostat. Magnetic field of 0.1-0.2 T was provided by a small permanent magnet placed under the substrate. The 100-GHz radiation from Gunn oscillator has been focused on the sample by the hemispheric Si-lens. The *I-V* characteristics have been measured with the low noise oscilloscope in the shielded room. Typically mesa had resistance at room temperature about 200 Ohm and critical current density at 4.2 K about 1 kA/cm$^2$.

Fig. 2a shows a set of the *I-V* characteristics of Bi-2212 mesa at 5K with an increase of microwave power. The first prominent feature is an appearance of a big voltage step with a sharp increase of current at nearly constant voltage. The voltage position of this step $V_{ff}$ increases proportionally to the square root of microwave power *P* (Fig. 3) while the current height of the step remains unchanged (Fig. 2a). This step resembles the Josephson flux-flow (JFF) step that appears in steady magnetic field oriented parallel to the layers [7] and is often referred to as microwave induced JFF step.

Another remarkable feature is an appearance of Shapiro-type steps on the *I-V* characteristics at voltages $nV_0$, where *n* is an integer and $V_0$ =10 mV corresponds to the Josephson relation $V_0 = Nh\nu/2e$ with $\nu$ the frequency and $N$=50 the number of IJJs in the mesa. Those steps are smaller than microwave induced JFF step. The heights of Shapiro steps, $\Delta I_n$, and of critical current $I_c$ oscillate with microwave power (Fig. 4), the steps with even *n* and *n*=0 oscillating in-phase with each other and out of phase with those of odd *n*. That behaviour is typical for Shapiro steps [8]. The position of the steps does not change with power as it is clearly seen from derivative picture (Fig. 2b). At high enough power we observed up to 10$^{th}$ harmonic of the main step. That corresponds to a frequency of 1 THz. A remarkable feature is that Shapiro steps appear only at voltages below the voltage position of microwave induced JFF step $V_{ff}$ (Figs. 2ab). The voltage position of microwave induced JFF step at high microwave power exceeds position of the 10$^{th}$ harmonic of Shapiro step. That implies that corresponding operating frequency of Josephson oscillations exceeds 1 THz. Microwave induced JFF step has been observed earlier in the intrinsic and conventional Josephson junctions [9-12] at zero magnetic field, however, that has not been accompanied by Shapiro steps and also has not been observed at so high bias voltages. Some authors considered the origin of microwave induced JFF step as being related with a parallel magnetic field induced by microwave current [9]. This idea hardly explains our data since that requires the presence of incredibly high parallel fields of several Tesla.

To explain our main results: synchronization of IJJ and terahertz operating frequencies under small perpendicular field of 0.1 T we consider the following qualitative model. As known, at low temperatures perpendicular magnetic field of about 0.1 T drives Bi-2212 system close to the transition into the vortex glass state [13]. This state is characterized by a disappearance of the hysteresis of the IV characteristics [14]. In that state pancake vortex lines become highly entangled as schematically shown in Fig. 5. That leads to the appearance in a system of Josephson vortices (JVs) and antivortices. Lorentz force induced by currents applied across the layers directly affects only JVs. Therefore, due to the weak attractive interaction between JVs and pancake vortex lines [15], the motion of JVs may be considered as a motion in a quasiperiodic washboard potential with a periodicity of pancake vortex lines [16]. The pancake vortex lines piercing the whole thickness of the mesa form aligned periodic potential in different IJJs and thus synchronize the motion of Josephson vortices in the whole mesa. For $H_\perp$=0.1 T the spacing between pancake vortex lines $\Lambda \sim (\Phi_0/H_\perp)^{1/2} \sim$ 0.15 μm. That provides terahertz washboard frequency for easily achievable velocity of JVs $10^6$-$10^7$ cm/s. Note for comparison that to achieve the same periodicity in Josephson vortex lattice by parallel field $H_{//}$ that requires about 10T ($H_{//}=\Phi_0/s\Lambda$, with s=1.5 nm the spacing between superconducting layers).

Two types of characteristic modes appear in this model. First one corresponds to the washboard frequency and equals to the *n* times of the external frequency where *n* is the number of periods which JV travels in one cycle of the microwave current. That is proportional to the amplitude of microwave current or $P^{1/2}$, so we attribute the observed microwave induced JFF step to the mode of this type. In a presence of DC current the motion of JV can be extended. Then mode locking appears when for one cycle of AC current JV is displaced in one direction on integer number *m* of periods. For instance, for *m*=1 it can move 2 periods forward and 1 backward. Those modes correspond to Shapiro steps of the $m^{th}$ order. From this consideration it is clear that *n* always exceeds *m*, i.e. frequency of Shapiro step harmonics is always less than washboard frequency. That is consistent with our experiment.

In a summary, we found that relatively weak magnetic field oriented across the layers effectively synchronizes intrinsic Josephson junctions with external microwaves and enables to reach terahertz operating frequencies in microwave induced Josephson flux-flow regime.

**Acknowledgement.** The work has been supported by Presidium of RAS Program No. 27 "Base researches of nanotechnologies and nanomaterials", RFBR-NNSF grant 06-02-39021-GFEN_a, RFBR grant 06-02-72551-CNRS_a, NSFC of China (grant Nos. 60571007 and 60610050), Korea Research Foundation grant KRF-2007-613-D00003, Korea Science and Engineering Foundation grant F01-2007-000-10202 and INTAS grant No. 05-1000008-7972.

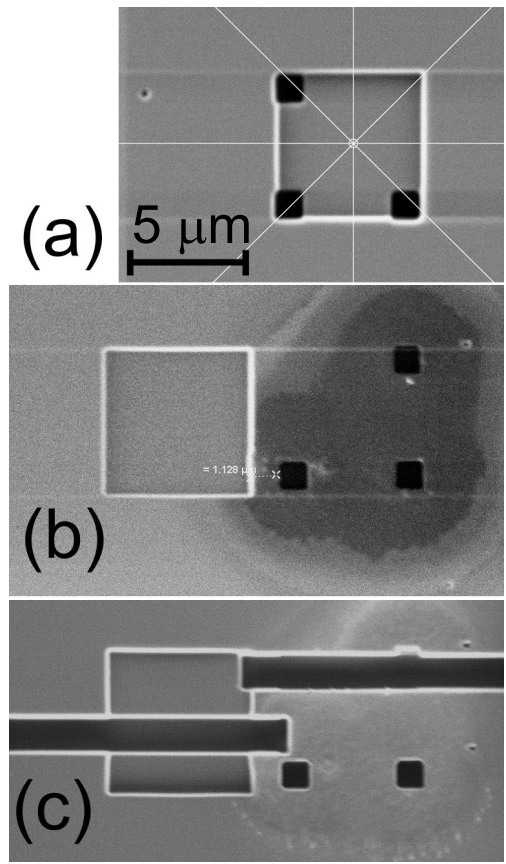

Fig. 1. The stages of fabrication of Bi-2212 mesa by double-sided processing of Bi-2212 single crystal whisker with focused ion beam (a-c).

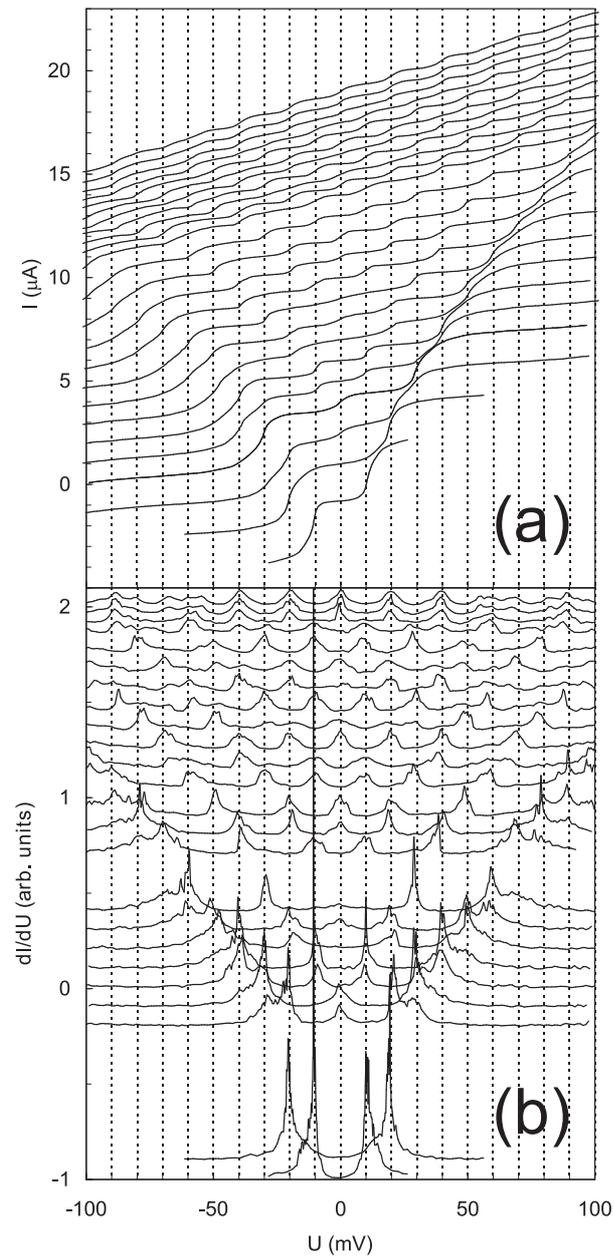

Fig. 2. Series of the *I-V* characteristics (a) and of the corresponding differential *I-V* characteristics (b) for the power of external 100 GHz radiation (bottom to top): -3.03; -1.15; 0.55; 2.23; 3.35; 4.41; 5.46; 6.36; 7.27; 8.17; 8.90; 9.63; 10.36; 10.93; 11.50; 12.07; 12.64; 12.88; 13.12; 13.44; 13.83; 14.23; 14.63; 15.03 dBm. Curves are offset vertically for clarity. T=5 K.

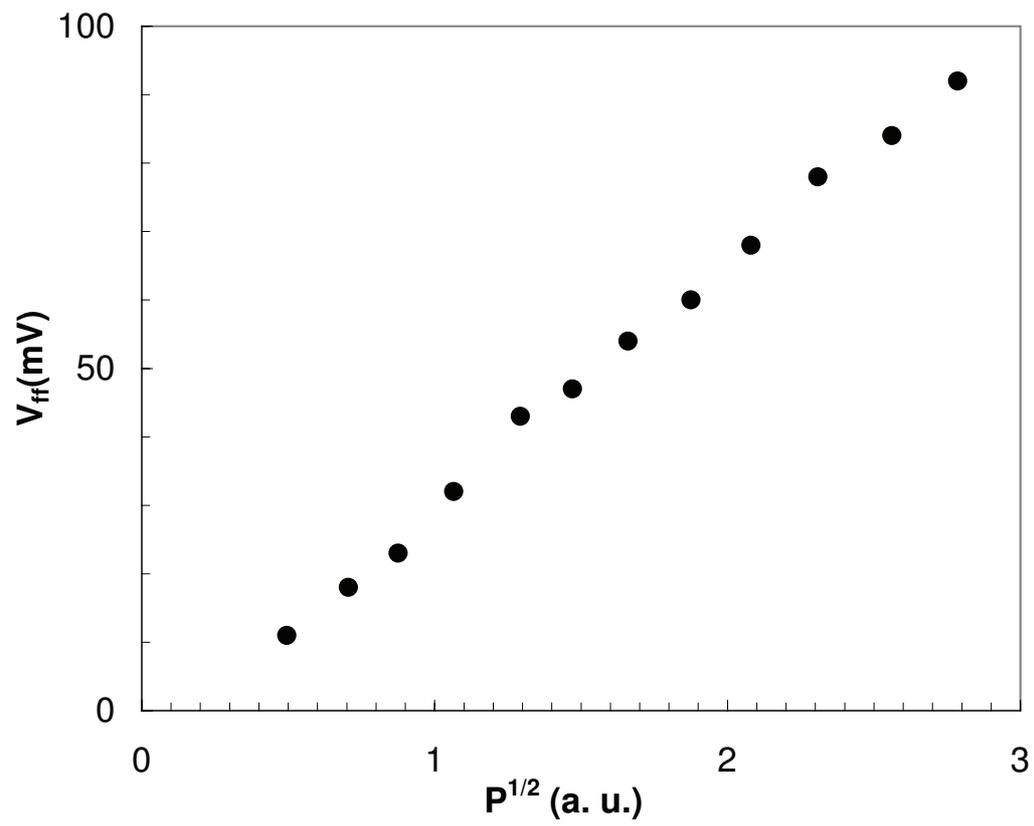

Fig. 3. Microwave induced flux-flow step position dependence on the power of external 100 GHz radiation. T=5 K.

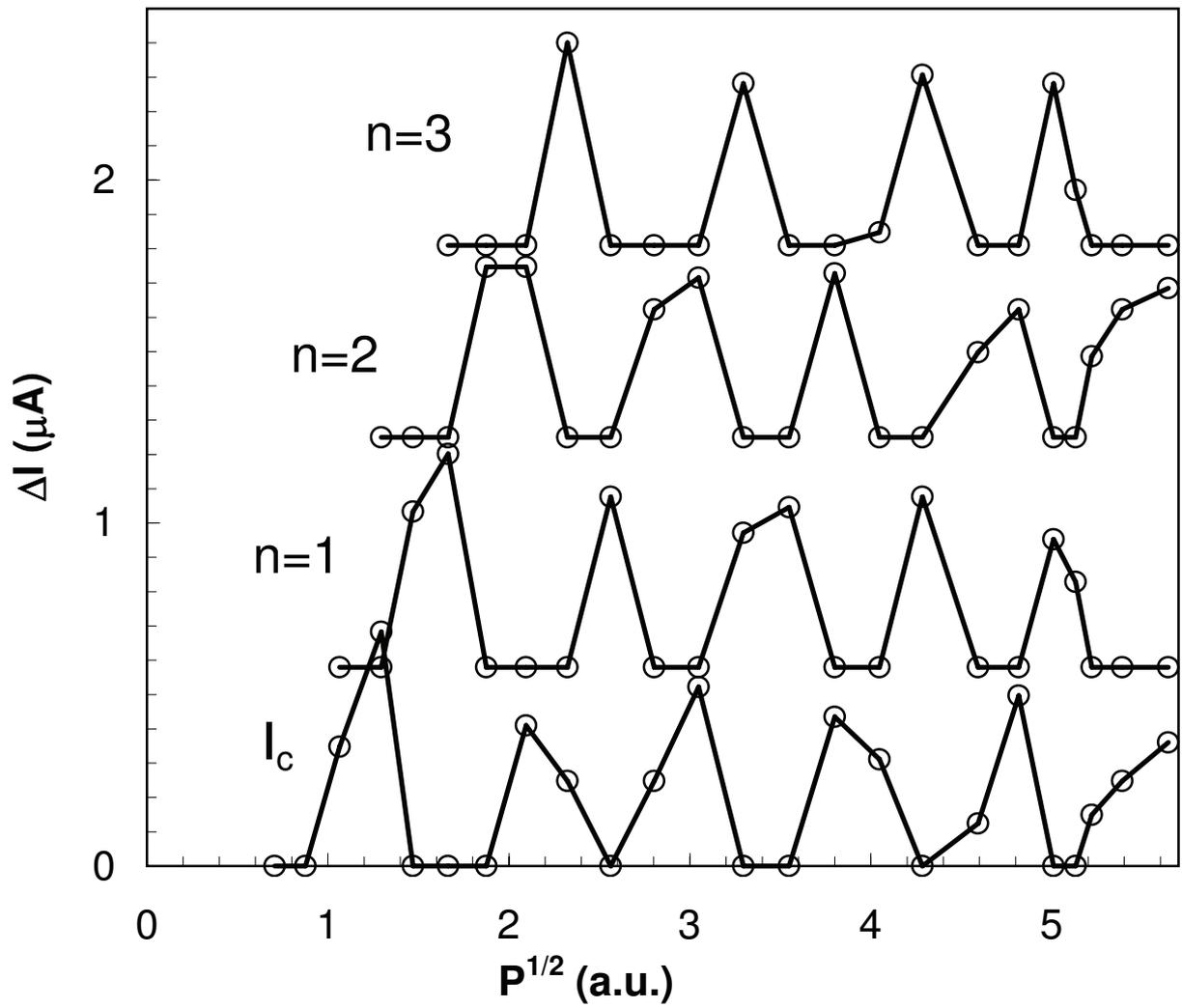

Fig. 4. Josephson critical current and three first Shapiro step height dependences on the power of external 100 GHz radiation. Curves are offset vertically for clarity. Lines are guides to eye. T=5 K.

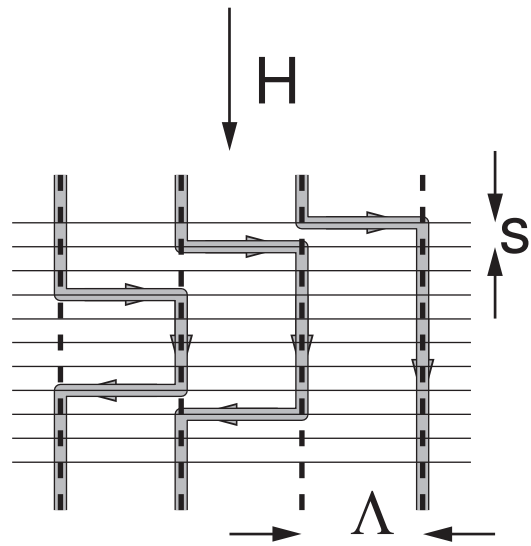

Fig. 5. Schematic view of flux lines in the vortex-glass state. For details see text.